\documentstyle[11pt,newpasp,twoside,psfig]{article}

\def\edcomment#1{\iffalse\marginpar{\raggedright\sl#1\/}\else\relax\fi}
\reversemarginpar

\begin{document}

\title{On the Merit of Observations Beyond the Cluster Core}
 \author{Kevin A. Pimbblet}
\affil{Department of Physics, University of Durham, South Road, Durham, DH1 3LE, United Kingdom}

\begin{abstract}
I discuss recent seminal work on the LARCS dataset: a panoramic study of rich 
clusters of galaxies at $z\sim0.12$.  The importance of observing
beyond the cluster core is illustrated by exploiting these 
data to examine colour gradients across the clusters.
\end{abstract}
\section{Introduction}
The Las Campanas / AAT Rich Cluster Survey (LARCS; O'Hely et al. 1998; 
Pimbblet et al. 2001a; 2001b; Pimbblet 2001) is a 
long-term project to study a statistically-reliable sample
of 21 of the most luminous X-ray clusters at intermediate redshifts
($z=0.07$--0.16) in the southern hemisphere.
The photometric imaging of these clusters comprises homogeneous, two degree
wide $B$ and $R$ band observations taken at Las Campanas Observatory.  
These data permit 
tracing of photometric variations in cluster members out to large 
radii (typically $\sim12$ Mpc at $z\sim0.12$).   
Galaxies selected from these data are being observed in an on-going
spectroscopic follow-up (Pimbblet 2001) with 
the 2dF spectrograph on the Anglo-Australian Telescope.

\section{Beyond the Core}
Recently, theoretical and observational attention has been turned
to the properties of galaxies beyond the cluster core, existing betwixt the 
infall regions and the virial radius of clusters 
(e.g. Balogh, Navarro, \& Morris 2000; 
Pimbblet et al. 2001b; Kodama et al. 2001).  
This region is crucial for understanding the
evolution of the galaxy population in clusters as they grow through
the accretion of field galaxies at $z\leq 1$, and hence to answer
important questions regarding the origin of such basic correlations as
the morphology-density relation seen at the present day (Dressler
1980).
At intermediate redshift most field galaxies are actively forming
stars. Yet, as these galaxies fall into clusters, their star formation
rate declines, eventually to zero (Poggianti et al. 1999).
The first clear demonstration of this transformation is reported
by Abraham et al. (1996) who show that the reddest spectroscopically-confirmed
cluster members, those lying on the colour-magnitude relation (CMR), in
the $z=0.23$ cluster Abell~2390, get progressively bluer as a function
of cluster radius out to $\sim5$ Mpc.  

Pimbblet et al.\ (2001b) analyze a sample of eleven LARCS clusters
and combine these to trace the CMR from 
the dense cluster core out to the low-density field.  
They report that a small blueward shift of 
$d(B-R) / dr = -0.022 \pm 0.004$ (or, equivalently in terms of 
local-density: $d(B-R) / d log_{10}(\Sigma) = -0.076 \pm 0.009$)
is present in the colours of the peak of the CMR, out to $\sim6$ 
Mpc; equivalent to $\Delta (B-R)\sim0.1$.

A statistical background correction technique, however, is 
used by Pimbblet et al.
(2001b) to define cluster membership, therefore the CMR radial blueing 
signal may be contaminated by background galaxies.  
Spectroscopic observations of the LARCS galaxies (Pimbblet 2001) are yielding 
membership information for individual galaxies needed to improve 
the signal-to-noise in the CMR at large radii where this contamination
becomes important.  Present results show a spectroscopically-confirmed 
blueward shift in the CMR's peak colour of $d(B-R) / dr = 
-0.017 \pm 0.005$ which is consistent
with the statistically-defined membership result. 
Pimbblet et al. (2001b) suggest that
these trends most likely reflect differences in the 
luminosity-weighted ages of the galaxies in different environments.  
If interpreted purely
as a difference in ages then the gradient observed within LARCS
suggests that the luminosity-weighted ages of the dominant galaxy
population within the CMR at 6\,Mpc from
the cluster core are some 3\,Gyrs younger than those 
residing in the core.

\section{Summary}
To conclude, 
wide-field observations are paramount to the understanding of cluster 
formation and evolution and to trace radial variations in the 
cluster population (i.e. the CMR is readily observed to 
beyond 8 Mpc in many LARCS clusters; Pimbblet 2001).
Although some recent work on a single high redshift cluster 
has already been undertaken by Kodama et al. (2001),  
much further work (photometric and spectroscopic) is urgently 
required to comparatively examine radial trends within a 
larger, well-defined sample of higher redshift clusters.
Such observations would allow the testing of the prediction
that these trends will be more strongly pronounced at higher
redshifts (Pimbblet et al. 2001b).


\begin{references}

\reference Abraham, R. G. et al. 1996, ApJ, 471, 694.

\reference Balogh, M. L., Navarro, J. F., \& Morris, S. L. 2000, \apj, 540, 113 

\reference Dressler, A. 1980, \apj, 236, 351  

\reference Kodama, T., Smail, I., Nakata, F., Okamura, S., Bower, R. G. 2001, \apj L, in press 

\reference O'Hely, E., Couch, W. J., Smail, I., Edge, A. C., Zabludoff, A. I. 1998, \pasp, 15, 3, 273

\reference Pimbblet, K. A., Smail, I., Edge, A. C., Couch, W. J., O'Hely, E., Zabludoff, A. I. 2001a, \mnras, 327, 588

\reference Pimbblet, K. A., Smail, I., Kodama, T., Couch, W. J., Edge, A. C., Zabludoff, A. I., O'Hely, E. 2001b, \mnras, in press, astro-ph/0111461

\reference Pimbblet, K. A. 2001, Ph.D. Thesis, Stellar Populations of X-Ray Luminous Clusters at $z=0.1$, University of Durham

\reference Poggianti, B. M. et al. 1999, \apj, 518, 576 


\end{references}
\end{document}